\documentclass[twocolumn]{aastex631}

\usepackage{amsmath}

\shorttitle{Recurrent Infrared Variability in NGC 2547 ID8}
\shortauthors{Chen et al.}

\begin{document}

\title{A Recurrent Infrared Modulation in the Warm Post-impact Debris Disk NGC 2547 ID8}

\author[0000-0001-7084-0484]{Xiaodian Chen}
\affiliation{CAS Key Laboratory of Optical Astronomy, National Astronomical Observatories, Chinese Academy of Sciences, Beijing 100101, China}
\affiliation{Institute for Frontiers in Astronomy and Astrophysics, Beijing Normal University, Beijing 102206, China}
\affiliation{School of Astronomy and Space Science, University of Chinese Academy of Sciences, Beijing 100049, China}

\author[0000-0003-4489-9794]{Shu Wang}
\affiliation{CAS Key Laboratory of Optical Astronomy, National Astronomical Observatories, Chinese Academy of Sciences, Beijing 100101, China}
\affiliation{School of Astronomy and Space Science, University of Chinese Academy of Sciences, Beijing 100049, China}

\author{Zhuochao Huang}
\affiliation{CAS Key Laboratory of Optical Astronomy, National Astronomical Observatories, Chinese Academy of Sciences, Beijing 100101, China}
\affiliation{School of Space and Earth Sciences, Beihang University, Beijing 102206, China}

\correspondingauthor{Xiaodian Chen}
\email{chenxiaodian@nao.cas.cn}
\correspondingauthor{Shu Wang}
\email{shuwang@nao.cas.cn}

\begin{abstract}
Extreme debris disks are signposts of violent rocky collisions, but it remains unclear whether their mid-infrared variability traces transient dust clouds or longer-lived post-impact reservoirs. We analyze a 19 yr mid-infrared record of the young solar analog NGC 2547 ID8, combining four spectra from 2007--2024 with 3--5 $\mu$m monitoring through 2026. The optically thin mineral surface remains compositionally stable: broad fits spanning multiple grain sizes give 550--577 K, and a common-basis decomposition with $T_{\rm surf}$ fixed at 560 K is dominated by 1 $\mu$m olivine/Mg-olivine and 0.1 $\mu$m forsterite. The W2-equivalent 4.5 $\mu$m excess, traced across six recurrent high states, is instead driven mainly by changes in the hot-base continuum component and defines a 956.8 d infrared recurrence timescale with cycle-to-cycle variations in timing and amplitude. This combination supports the interpretation of ID8 as an early post-impact stage in which material on eccentric orbits repeatedly interacts with a warm rocky reservoir, preserving the impact's mineral fingerprint while the interaction geometry evolves. Material on eccentric orbits can persist for decades, and possibly longer, while interacting with debris disks. Longer-baseline, higher-cadence infrared monitoring of ID8 will further constrain the stability of its recurrence timescale and the detailed evolution of its infrared variability.
\end{abstract}

\keywords{Debris disks (363) --- Circumstellar dust (236) --- Planet formation (1241) --- Infrared excess (788) --- Time series analysis (1916)}

\section{Introduction}

Giant impacts are expected during the final growth and later reassembly of rocky planetary systems, when embryos and planetesimals collide after the gas disk has dispersed \citep{Chambers2001,Asphaug2006,Wyatt2008,Hughes2018}. Their observable debris may include condensate clouds, optically thin silicate grains, bound fragments, and eccentric streams whose collisional evolution can maintain warm infrared excesses beyond the immediate aftermath of a single impact \citep{JacksonWyatt2012,Genda2015,WyattJackson2016,Watt2021,Lewis2023,Watt2024}. Extreme debris disks (EDDs) are therefore useful laboratories for post-impact evolution because they combine high fractional infrared luminosities, strong solid-state features, and multi-year variability \citep{Meng2015,Su2019}.

Semi-regular infrared variability has been observed in a few EDDs. NGC 2547 ID8 showed roughly 10--35 d modulations, while P1121 retained a $16.7\pm1.5$ d modulation across five years, interpreted in terms of the orbital evolution and changing projected area of impact-produced dust \citep{Su2019}. TYC 4209-1322-1 exhibited approximately 39 d modulations during an 877 d \textit{Spitzer} campaign in 2017--2019 \citep{Moor2022}. Combining WISE and SPHEREx now extends the infrared photometric baseline to approximately 16 yr (2010--2026), enabling tests of multi-cycle year-scale recurrence.

Observed outcomes now range from rapidly disappearing warm dust and transiting collision clouds to repeated planetesimal collisions and candidate systems with evolving debris from catastrophic impacts \citep{Melis2012,Kenworthy2023,Kalas2026,Tzanidakis2026}. This diversity makes it necessary to distinguish a transient dust event from a bound post-impact system that can continue supplying observable grains.

Recent JWST/MIRI spectroscopy has sharpened this problem. The EDD sample studied by \citet{Su2026EDD} showed that warm EDD spectra commonly contain submicron grains in optically thin emission, thermally processed silicates, indices of silica- or forsterite-rich dust, and stochastic 3--5 $\mu$m variability. In parallel, the fading warm disk HD 15407A shows that a rocky mineral fingerprint can remain stable while the mid-infrared excess declines over decade timescales, consistent with an evolved post-impact reservoir dominated by large 5 $\mu$m grains \citep{Wang2026HD15407A}. These results motivate a direct test that combines spectroscopy with time-domain monitoring: does an EDD's infrared variability accompany a changing mineral inventory, or does it trace the interaction of material on eccentric orbits with an otherwise stable rocky reservoir?

NGC 2547 ID8 (hereafter ID8, at R.A. $=08^{\rm h}09^{\rm m}02\fs51$ and decl. $=-48^\circ58'17\farcs3$, J2000) is a benchmark system for this test. It is a $\simeq35$ Myr-old solar analog with a bright warm excess and strong infrared variability \citep{Gorlova2007,Meng2012,Meng2014,Meng2015,Su2019}. Earlier work attributed its rapidly varying warm emission to impact-produced dust clouds, with roughly 10--35 d modulations tracing their orbital geometry and dispersal \citep{Meng2014,Su2019}. The events identified in 2012 and 2014 were interpreted as distinct impacts, potentially linked by returning fragments \citep{Su2019}. Recurrent year-scale behavior and its relationship to the persistent mineral emission have not been established, limiting further constraints on the formation mechanism and evolution of this system. Here we combine four \textit{Spitzer} and JWST mid-infrared spectra spanning 17 yr (2007--2024) with 3--5 $\mu$m photometry to show that ID8 separates two observables: a stable small-grain silicate surface and a recurrent, variable hot-base continuum component. Section 2 describes the data and modeling approach. Section 3 presents the stability of the mineral surface and the variability of the hot-base continuum component. Section 4 characterizes the W2-equivalent 4.5 $\mu$m recurrence timescale. Section 5 places ID8 in the broader JWST EDD population, and Section 6 discusses the physical interpretation and implications for post-impact evolution.

\section{Data and Model}

We use the low-resolution \textit{Spitzer}/IRS spectrum of ID8 observed on 2007 June 16 in GTO program 40227 (PI: G. Rieke, AOR 21755136). The spectrum was obtained from the CASSISjuice LRv7 \texttt{LRTaperedPLS} branch, which combines tapered-column extraction with point-like-source flux calibration \citep{Houck2004,Lebouteiller2011,Lebouteiller2015}. For the cross-instrument comparison, we retain a direct 5.3--26.9 $\mu$m cut of the native flux-density and uncertainty samples at $R\simeq60$--127, without rescaling or rebinning, following the product-preserving treatment used for HD 15407A \citep{Wang2026HD15407A}. We also use three JWST/MIRI Medium Resolution Spectrometer (MRS) spectra obtained in GTO program 1206 (PI: G. Rieke), specifically Observations 1, 8, and 10, hereafter o1, o8, and o10 \citep{Rieke2015,Wells2015,Wright2023}. For JWST, we adopt the current Level 3 x1d products from MAST. They were calibrated at the observation level with pipeline version 1.20.2 and CRDS context \texttt{jwst\_1464.pmap}, using reference file \texttt{jwst\_miri\_extract1d\_0004.asdf} for point-source extraction. Each observation used four dithers in each of the short, medium, and long MRS grating settings on the MIRIFUSHORT and MIRIFULONG detectors, giving 24 \texttt{UNCAL} exposures and 12 Level 3 channel--band spectra at $R\simeq1500$--3500. We independently reprocessed all three observations from their \texttt{UNCAL} exposures through the default \texttt{Detector1Pipeline}, \texttt{Spec2Pipeline}, and \texttt{Spec3Pipeline} stages using pipeline version 2.0.1 and CRDS context \texttt{jwst\_1535.pmap}. Across the 11 channel--band segments used over 4.9--22.0 $\mu$m, the median ratios of independently reduced fluxes to archive fluxes are 1.0008, 1.0008, and 1.0012 for o1, o8, and o10, respectively, and the maximum segment offsets are 0.23\%, 0.41\%, and 0.54\%. Given these small differences, we retain the MAST Level 3 products and concatenate their native wavelength grids without multiplicative overlap rescaling.

We use the Dust Continuum Kit (DuCK), the dust-continuum part of DuCKLinG (Dust Continuum Kit with Line emission from Gas), introduced by \citet{Kaeufer2024}. Its framework combines thermal continuum emission with a library of dust absorption-efficiency curves, supports nested sampling, and can be extended with additional dust components. DuCK has been applied to the JWST/MIRI and \textit{Spitzer}/IRS spectra of the protoplanetary disk PDS 70 \citep{Jang2024} and adapted to the debris disk HD 15407A \citep{Wang2026HD15407A}. For ID8, we use the dust-only model described below.

\begin{figure*}[!t]
\centering
\includegraphics[width=0.98\textwidth]{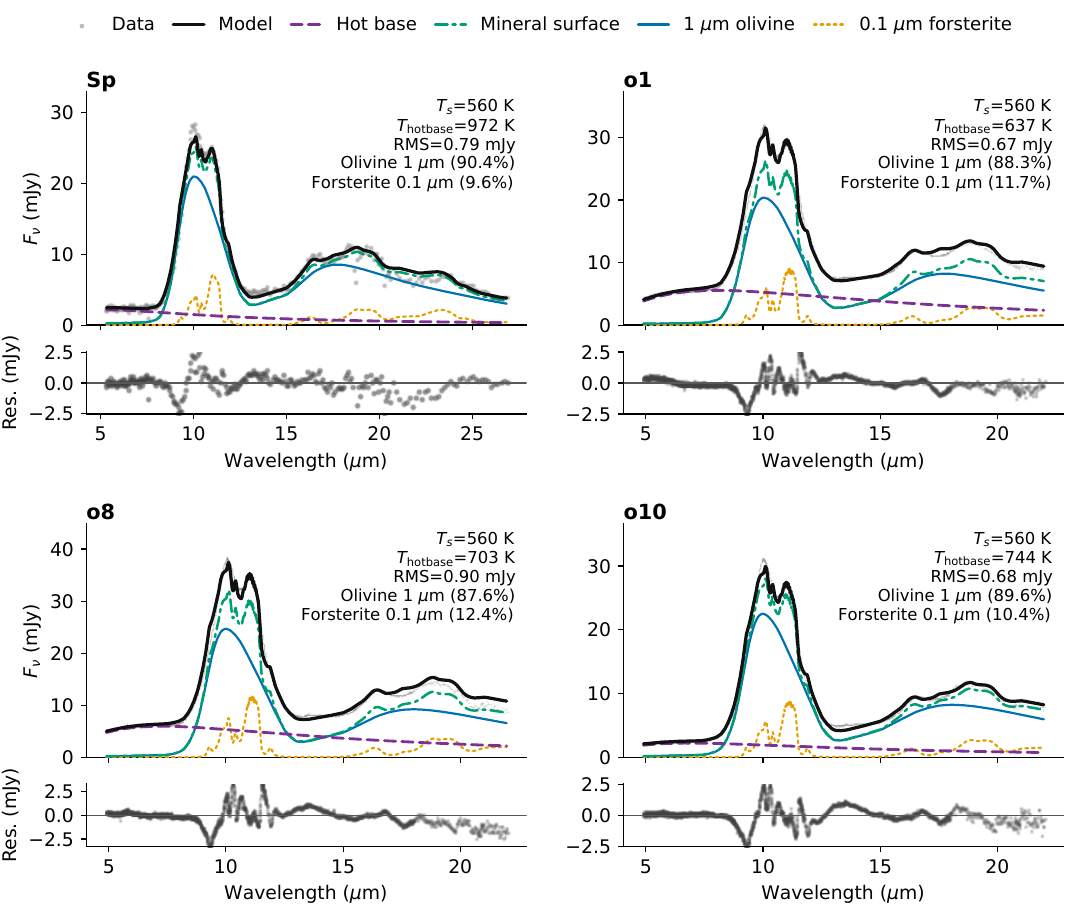}
\caption{Dust-emission decomposition of ID8 at four epochs. \textit{Spitzer} (Sp) and JWST/MIRI (o1, o8, and o10) spectra are fitted with the two-temperature dust model in Equation (1) after subtraction of the stellar photosphere. The mineral surface is fixed at 560 K, representative of the 550--577 K range from the broad multi-size fits. Gray points show the data. The black curves show the total model, while the purple dashed, green dash-dotted, blue, and orange dotted curves show the hot-base continuum component, total optically thin mineral emission, combined 1 $\mu$m olivine/Mg-olivine emission, and combined 0.1 $\mu$m forsterite emission, respectively. The grouped mineral fractions remain stable, while the hot-base continuum component varies strongly.}
\label{fig:spectra}
\end{figure*}

\begin{deluxetable*}{lcccccccc}
\tabletypesize{\scriptsize}
\tablecaption{Hot-base and mineral-surface parameters across the four spectra\label{tab:mineral}}
\tablehead{
\colhead{Epoch (UTC)} &
\colhead{MJD} &
\colhead{$T_{\rm surf}$} &
\colhead{$T_{\rm hot}$} &
\colhead{$C_{\rm hot}$} &
\colhead{$F_{\rm W2}^{\rm exc}$} &
\colhead{$M_{\rm surf}$} &
\colhead{Olivine/Mg-olivine} &
\colhead{Forsterite} \\
\colhead{} &
\colhead{} &
\colhead{(K)} &
\colhead{(K)} &
\colhead{($10^{-18}\,{\rm sr}$)} &
\colhead{(mJy)} &
\colhead{($10^{-7}\,M_\oplus$)} &
\colhead{(mass \%)} &
\colhead{(mass \%)}
}
\startdata
Sp (2007-06-16) & 54267.0 & 560 (fixed) & $972_{-44}^{+48}$ & $0.127_{-0.014}^{+0.016}$ & $2.427_{-0.087}^{+0.088}$ & $5.8\pm0.0$ & $90.4\pm0.1$ & $9.6\pm0.1$ \\
o1 (2022-12-28) & 59941.9 & 560 (fixed) & $637.0\pm0.2$ & $1.136\pm0.001$ & $3.721_{-0.001}^{+0.002}$ & $5.8\pm0.0$ & $88.3\pm0.0$ & $11.7\pm0.0$ \\
o8 (2023-05-15) & 60079.5 & 560 (fixed) & $703.1\pm0.2$ & $0.910\pm0.001$ & $4.649\pm0.002$ & $7.1\pm0.0$ & $87.6\pm0.0$ & $12.4\pm0.0$ \\
o10 (2024-01-07) & 60316.7 & 560 (fixed) & $743.9_{-0.6}^{+0.7}$ & $0.289\pm0.001$ & $2.000\pm0.002$ & $6.4\pm0.0$ & $89.6\pm0.0$ & $10.4\pm0.0$
\enddata
\tablecomments{$F_{\rm W2}^{\rm exc}$ is the W2-equivalent flux after bandpass convolution and stellar subtraction, and $C_{\rm hot}$ is the hot-base emission coefficient. $M_{\rm surf}$ is the model-dependent emitting mass in the fitted optically thin basis of 0.1--1 $\mu$m grains.}
\end{deluxetable*}

Following the photospheric treatment of \citet{Su2019}, the fixed stellar component in all four DuCK fits was taken from the solar-metallicity Castelli--Kurucz ATLAS9 grid of stellar-atmosphere spectra, hereafter CK04 \citep{CastelliKurucz2003}, using $T_\star=5500$ K and $\log g=4.5$. Su et al. used $d=361$ pc and $R_\star\simeq0.95\,R_\odot$. We updated the distance to the Gaia DR3 value of 352 pc and adopted $R_\star=0.925\,R_\odot$ to preserve the angular-radius scale $R_\star/d$ \citep{GaiaDR3,Creevey2023,Su2019}. Absolute bandpass convolution gives photospheric fluxes of 5.612 mJy in \textit{Spitzer} [4.5] and 5.292 mJy in W2. These choices change the corresponding photospheric fluxes by less than 0.5\% relative to the previous stellar scale. The differences are negligible for the present analysis. The photometric record includes \textit{Spitzer} [4.5] measurements \citep{Fazio2004,Su2019}, AllWISE and NEOWISE W1/W2 photometry \citep{Wright2010,Mainzer2011,Mainzer2014}, four W2-equivalent points synthesized from the spectra described above, and W2-equivalent fluxes from three SPHEREx epochs \citep{Crill2020}. We used the Quick Release 2 (QR2) Level-2 calibrated spectral images \citep{SPHERExQR2}, covering 2025 April--May, 2025 November--December, and 2026 April--May. We subtracted the corresponding bandpass photosphere to place all measurements on a common excess scale. Hereafter, 4.5 $\mu$m excess denotes this stellar-subtracted W2-equivalent scale unless a specific instrumental band is named. Because ID8 lies in a crowded cluster field, each 0.17$^\circ$ SPHEREx Level-2 cutout was fitted with oversampled PSF templates integrated over detector pixels. The templates were fixed at the position of ID8 and those of 11 neighboring AllWISE sources within 60 arcsec, and the fit included terms for zodiacal emission and the local background. The extracted 3.89--5.08 $\mu$m ID8 spectra were extrapolated across the weak red tail of W2 with a fitted thermal continuum and convolved with the full response. This gives excesses of $3.182\pm0.048$, $3.526\pm0.076$, and $4.012\pm0.099$ mJy at median MJDs 60798.262, 60999.882, and 61163.634, respectively. The uncertainties combine bootstrap scatter from the PSF extraction with uncertainty from extrapolating the thermal tail.

We represent the stellar-subtracted spectra with the dust-only DuCK model,
\begin{equation}
F_{\lambda}^{\rm exc} =
C_{\rm hot}B_\lambda(T_{\rm hot})+
B_\lambda(T_{\rm surf})\sum_i C_i\kappa_i(\lambda),
\end{equation}
where $F_{\lambda}^{\rm exc}$ is the stellar-subtracted excess, the first term is a smooth hot-base continuum component, and the second term is emission from an optically thin mineral surface. For each sampled $T_{\rm hot}$, $C_{\rm hot}$ and the mineral amplitudes $C_i$ were solved by non-negative least squares \citep{LawsonHanson1974}. As in the no-rim framework of \citet{Wang2026HD15407A}, exploratory fits with a normal inner rim and broad surface/midplane temperature ranges did not require a separate rim component. We therefore adopted a two-temperature, isothermal-component dust model. Because ID8 is a gas-poor debris disk, molecular lines were not included.

\section{A Stable Mineral Surface}

The four spectra are well reproduced by a common mineral basis (Figure~\ref{fig:spectra}). A broad initial screen over eight mineral families and 0.1, 1, and 5 $\mu$m grains placed almost all of the optically thin dust in olivine/Mg-olivine and forsterite, with every other mineral family contributing less than 5\% by mass, and gave $T_{\rm surf}=550$--577 K across the four epochs. We therefore fixed $T_{\rm surf}=560$ K and fitted every epoch with the same compact basis containing the mineral components that exceeded the 5\% screening threshold: 1 $\mu$m olivine, 1 $\mu$m Mg-olivine, and 0.1 $\mu$m forsterite represented by distribution-of-hollow-spheres (DHS) and Gaussian-random-field (GRF) opacities \citep{Dorschner1995,Fabian2001}. This controlled reduction isolates the dominant sources of cross-epoch spectral variability but can omit minor, potentially real minerals and small changes in the surface temperature. A matched fit that removes both forsterite opacities increases the 8--13 $\mu$m residual rms by 0.9--1.5 mJy and is disfavored by a minimum $\Delta{\rm BIC}$ of 1270 across the four epochs, supporting forsterite within the adopted opacity basis.

Table~\ref{tab:mineral} summarizes the fitted parameters and their formal uncertainties. The resulting optically thin surface has a stable mineral partition. The model-dependent emitting mass in visible grains ($a\leq1\,\mu$m) remains within $5.8$--$7.1\times10^{-7}\,M_\oplus$. The combined 1 $\mu$m olivine/Mg-olivine component contributes 87.6--90.4\% by mass, and the combined 0.1 $\mu$m forsterite component contributes 9.6--12.4\%. In contrast, the hot-base continuum component changes strongly. The W2-equivalent excess synthesized from the spectra ranges from 2.000 to 4.649 mJy, and the fitted hot-base color temperature spans 637--972 K. The two high-flux epochs, o1 and o8, require $C_{\rm hot}=(0.910$--$1.136)\times10^{-18}$ sr, compared with $(0.127$--$0.289)\times10^{-18}$ sr for Sp and o10 (Table~\ref{tab:mineral}). Changes in the hot-base emission coefficient, $C_{\rm hot}$, dominate the variation in the hot-base continuum component, while temperature variations provide only a secondary compensation. Therefore, the stable silicate fingerprint records a persistent small-grain surface, while the changing 3--5 $\mu$m emission is carried by the hot-base continuum component.

\section{Recurrent Infrared Variability}

\begin{figure*}[!t]
\centering
\includegraphics[width=\textwidth]{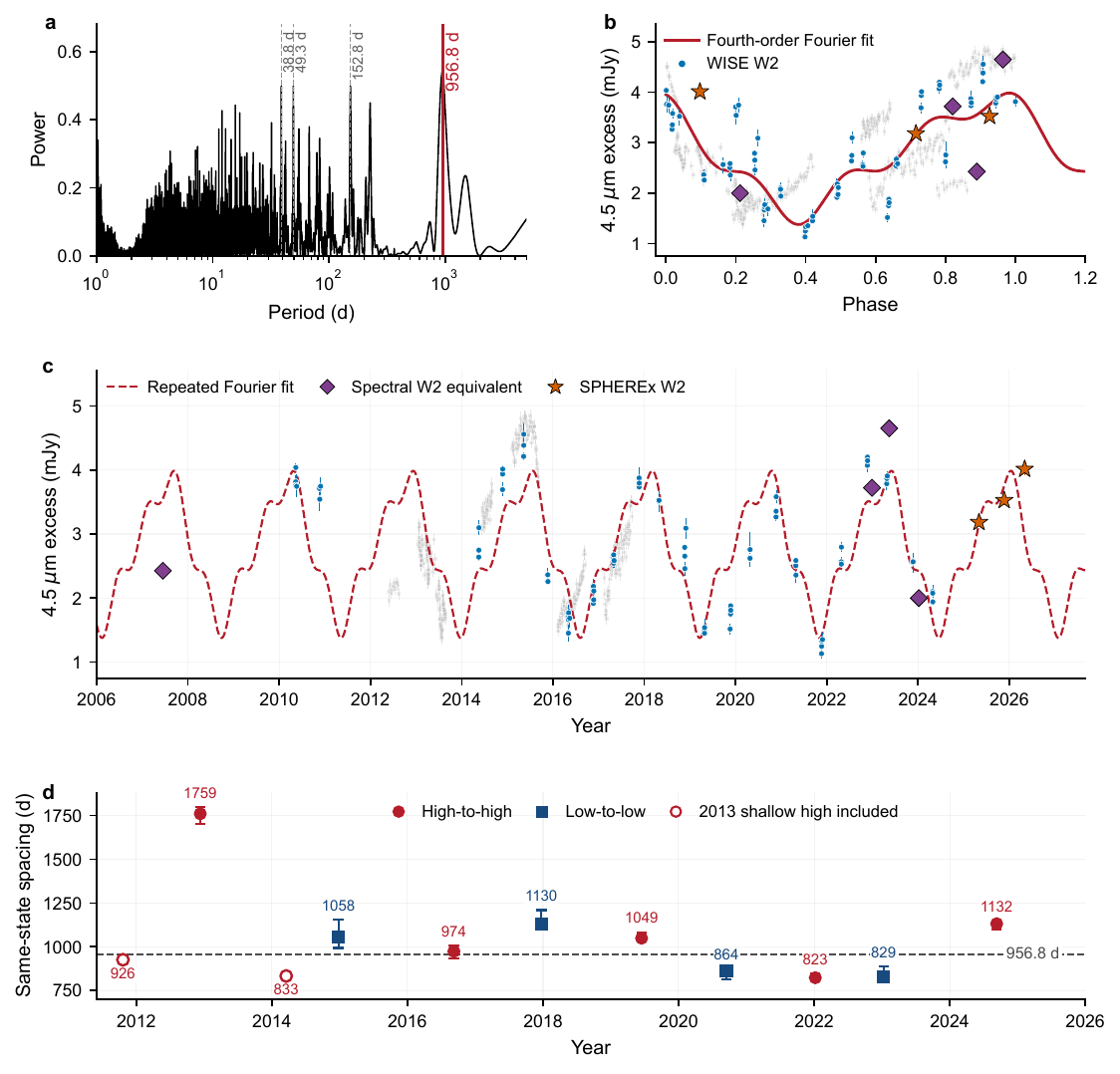}
\caption{Recurrent infrared variability and its cycle-to-cycle changes. (a) Lomb--Scargle periodogram of the combined WISE W1/W2 light curve. The red line marks $P_{\rm rec}=956.8$ d and gray dashed lines mark the principal aliases. (b) W2-equivalent excesses on the common CK04 photospheric scale, folded at the mean recurrence timescale. The red curve is a weighted fourth-order Fourier fit to the complete data set. Gray points, blue circles, purple diamonds, and orange stars denote \textit{Spitzer} [4.5], WISE W2, spectroscopic W2-equivalent, and SPHEREx W2-equivalent measurements, respectively. (c) Full 4.5 $\mu$m excess record as a function of time. The dashed red curve is the same fourth-order Fourier fit repeated at the recurrence timescale. (d) Spacings between successive high states or successive low states, plotted at the temporal midpoint of each pair. Open red circles show the alternative spacings obtained when the shallow 2013 \textit{Spitzer} maximum is treated as an additional high state. The spacings vary non-monotonically around the mean recurrence timescale marked by the dashed line.}
\label{fig:clock}
\end{figure*}

The long-baseline W2-equivalent 4.5 $\mu$m excess reveals recurrent high and low states (Figure~\ref{fig:clock}). For the periodogram in Figure~\ref{fig:clock}a, daily-binned W1 and W2 measurements were separately demeaned and combined in a Lomb--Scargle search \citep{Lomb1976,Scargle1982}. The strongest peak defines a mean recurrence timescale of $P_{\rm rec}=956.8$ d. W2 alone peaks at the same timescale, while the shorter peaks near 38.8, 49.3, and 152.8 d are sampling aliases of the recurrence rather than independent periodicities.

To compare the heterogeneous measurements in time, we constructed a common 4.5 $\mu$m timing set with 509 W2-equivalent points. At fixed $P_{\rm rec}$, we folded the complete data set using $\phi=[(t-t_0)/P_{\rm rec}]\bmod 1$, where $t_0$ is an arbitrary phase-zero epoch, and fitted the fourth-order Fourier function $F(\phi)=a_0+\sum_{k=1}^{4}[a_k\sin(2\pi k\phi)+b_k\cos(2\pi k\phi)]$ by weighted linear least squares. To prevent one data source from dominating this template, uncertainty scale factors of 1, 3, and 9 were applied to the spectra/SPHEREx, WISE, and \textit{Spitzer} measurements, respectively. The four spectroscopic and three SPHEREx points broadly follow this template, while the largest departures occur in the earlier \textit{Spitzer} record (Figure~\ref{fig:clock}b,c).

To examine whether the recurrence timescale varies, we measured same-state spacings from flux-contrast-weighted high and low states. The binning, thresholds, centroid definitions, and error-bar construction are described in Appendix~\ref{app:state}. The six high-state centers give spacings of 1759, 974, 1049, 823, and 1132 d, while the five low-state centers give spacings of 1058, 1130, 864, and 829 d. Before the latest high state traced by SPHEREx, the spacings show a broad decrease from the longest early value through roughly 1000--1130 d to about 830--860 d, together with cycle-to-cycle fluctuations. The three SPHEREx epochs, however, show a clear rise in the 4.5 $\mu$m excess and place the latest high-state center in early 2026, giving a spacing of 1132 d. The longest spacing of 1759 d depends on whether the shallow 2013 \textit{Spitzer} [4.5] maximum is counted as a high state. It lies below our adopted 3.35 mJy threshold and is therefore excluded from the six high-state centers. Including it as an additional high state divides the 1759 d spacing into intervals of 926 and 833 d (open red circles in Figure~\ref{fig:clock}d). The same-state spacings therefore vary non-monotonically around the mean recurrence timescale. The high-state timing and infrared amplitude differ from cycle to cycle. Such variability can arise if material on eccentric orbits repeatedly reaches the warm emitting region while the emitting area, optical depth, or dust-production rate differs between passages.

\section{ID8 Among Warm EDDs}

\begin{figure*}[t]
\centering
\includegraphics[width=0.78\textwidth]{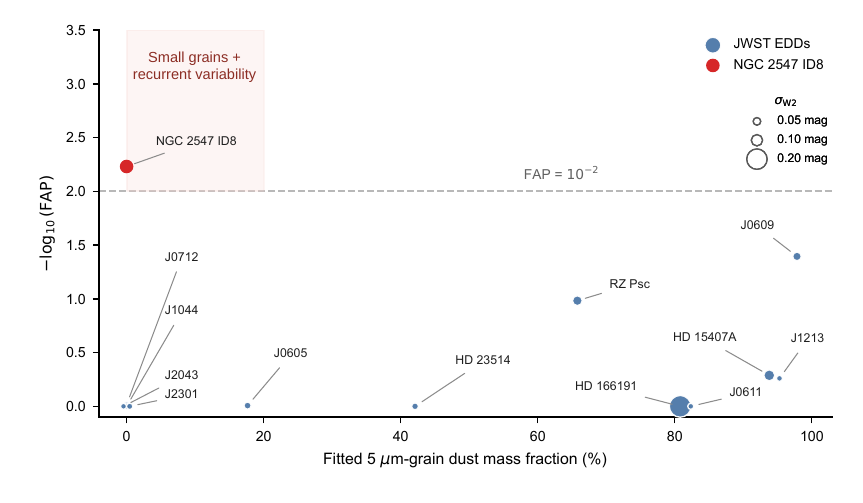}
\caption{ID8 in the population of warm EDDs observed with JWST. The fitted dust-mass fraction in 5 $\mu$m grains is compared with the periodogram significance derived from WISE W2 epochs, $-\log_{10}({\rm FAP})$. Symbol area encodes the W2 multi-epoch scatter. ID8 is the only system that combines a fitted surface dominated by small grains with W2 FAP $<10^{-2}$. Its $\sigma_{\rm W2}>0.10$ mag also shows pronounced infrared variability.}
\label{fig:population}
\end{figure*}

The stable mineral surface and recurrent W2 variability place ID8 in a distinct part of the current population of warm EDDs observed with JWST (Figure~\ref{fig:population}). We compiled JWST spectra of 16 systems for DuCK mineral decomposition, of which 15 also have WISE W2 variability diagnostics. After excluding J0925 and HD 131835, whose disk types differ from warm EDDs, 13 warm debris-disk systems, including ID8, remain for comparison. The JWST spectra of the 13 \mbox{systems} shown in Figure~\ref{fig:population} were obtained in programs 1206 (PI: G. Rieke) and 3189 (PI: K. Su). We applied the spectral-fitting method used for ID8 to obtain simplified mineral decompositions from the JWST spectra. Here we focus only on dust grain sizes. The detailed mineralogical analysis will be presented by Chen et al. (in preparation). For the photometric comparison, we applied a similar W2 periodogram procedure to ID8 and the other systems. W2 periodograms were calculated from independent WISE epoch medians over $P_{\min}=\max(300\,{\rm d},2\Delta t_{\rm epoch})$ to $P_{\max}=2000$ d, where $\Delta t_{\rm epoch}$ is the median time separation between adjacent independent WISE epochs. Significance was reported using the Baluev false-alarm probability (FAP) \citep{Baluev2008}. ID8 is the only system that combines a fitted surface dominated by small grains with W2 FAP below $10^{-2}$. Its W2 multi-epoch scatter also exceeds 0.10 mag, showing pronounced infrared variability.

ID8 may therefore represent an early post-impact stage. Its optically thin surface is dominated by 0.1--1 $\mu$m grains, and the stable spectra over a 17 yr baseline constrain a persistent size--composition distribution. For the smallest grains, radiation pressure reduces the effective stellar gravity and can place newly produced dust on unbound trajectories \citep{Burns1979}. The ratio of radiation pressure to stellar gravity is $\beta\equiv F_{\rm rad}/F_{\rm grav}=3L_\star Q_{\rm pr}/(16\pi GM_\star c\rho a)$. Here $L_\star$ and $M_\star$ are the stellar luminosity and mass. $Q_{\rm pr}$ is the radiation-pressure efficiency averaged over the stellar spectrum, $a$ is the grain radius, and $\rho$ is the material density. $G$ and $c$ are the gravitational constant and speed of light. Following standard debris-disk dynamics, a grain released from a circular parent-body orbit becomes unbound for $\beta>0.5$ \citep{Burns1979,Wyatt2008}. For $L_\star=0.705\,L_\odot$, $M_\star=0.953\,M_\odot$, and $\rho=3\,{\rm g\,cm^{-3}}$, this gives a nominal blowout size $a_{\rm blow}=0.28Q_{\rm pr}\,\mu$m.

For the fitted 0.1 $\mu$m forsterite, $\beta=1.42Q_{\rm pr}$, so it becomes unbound when the radiation-pressure efficiency averaged over the stellar spectrum satisfies $Q_{\rm pr}>0.35$. In the $Q_{\rm pr}=1$ limit, a grain released near the 560 K blackbody-equivalent scale of 0.21 au is blown beyond 1 au in 21.3 d. If instead $Q_{\rm pr}\leq0.35$, the grain remains bound and Poynting--Robertson drag acts on $t_{\rm PR}=4\pi\rho a c^2r^2/(3L_\star Q_{\rm pr})=13.1/Q_{\rm pr}$ yr, or at least about 37 yr. Here $r=0.21$ au and $t_{\rm PR}$ is the drag timescale under the circular, bound-orbit approximation. The dominant 1 $\mu$m olivine/Mg-olivine component has a longer corresponding removal time, $t_{\rm PR}=131/Q_{\rm pr}$ yr.

The fitted 0.1 $\mu$m forsterite component contains $5.5\times10^{-8}$--$8.8\times10^{-8}\,M_\oplus$. In the $Q_{\rm pr}=1$ limit, the 21.3 d travel time from 0.21 au to 1 au gives a replenishment scale of $1.6\times10^{-5}$--$2.6\times10^{-5}\,M_\oplus$ over the 17 yr spectral baseline, equivalent to a rocky body about 400--460 km in diameter for $\rho=3\,{\rm g\,cm^{-3}}$. This processing requirement supports the interpretation that ID8 is still in an early post-impact stage: the stable mineral surface is more naturally explained as a collisional cascade fed by a reservoir of larger parent bodies.

\citet{Su2026EDD} examined the population properties of EDDs, showing that they are characterized by optically thin submicron grains, thermally altered silicates, and infrared variability. In that context, the comparison in Figure~\ref{fig:population} suggests that, relative to ID8, other EDDs with fitted surfaces rich in small grains may contain material on eccentric orbits that produces variability too weak for WISE photometry to detect, or recurrence timescales too short for the WISE cadence to resolve. Systems dominated by 5 $\mu$m grains, such as HD 15407A, instead resemble more reprocessed, later-stage reservoirs in which the feeding geometry from a recent giant impact is no longer evident \citep{Wang2026HD15407A}.

\section{Discussion and Conclusions}

We considered several possible explanations for the joint spectral and photometric behavior of ID8. Large changes intrinsic to the optically thin mineral surface are not supported because they would be expected to alter the 8--13 $\mu$m silicate profile together with the 3--5 $\mu$m continuum. A sequence of unrelated impacts is also unlikely: they would have to preserve the same small-grain mineral inventory while producing recurrent year-scale 4.5 $\mu$m variability. A circular disk structure cannot satisfy the dust temperature and recurrence timescale simultaneously. Material at the 560 K blackbody-equivalent radius of about 0.21 au has a circular period of only about 36 d, whereas a circular belt with a 956.8 d period would be too distant to dominate the 3--5 $\mu$m emission. These constraints favor material on eccentric orbits that repeatedly reaches the warm region.

Earlier \textit{Spitzer} monitoring of ID8 resolved additional variability on timescales of approximately 10--35 d. \citet{Su2019} argued that these modulations require at least two violent impacts involving large asteroid-sized bodies. They further proposed that the events need not be independent: fragments produced by an initial grazing or hit-and-run collision could return and trigger a subsequent impact. The modulation attributed to the later impact weakened after about ten orbits, consistent with Keplerian shear erasing the non-axisymmetric structure of an individual impact cloud. These short-lived signals differ from the year-scale infrared recurrence identified here. The two behaviors may nevertheless probe complementary scales of the same post-impact evolution. Short-period modulations trace local structure near a collision point, whereas the persistent year-scale recurrence may trace bound material on eccentric orbits that repeatedly interacts with the warm region.

In this eccentric-orbit interpretation, the 560 K optically thin surface is a quasi-steady mineral layer maintained by a collisional cascade, while the hot-base continuum component records variable thermal emission associated with repeated passages of material on eccentric orbits through the warm region. The fitted $10^{-7}\,M_\oplus$ mass in the small grains that produce the surface emission is therefore only the visible component of the processed material. The required mass budget must be larger than this surface mass. One estimate comes from the hot-base continuum component. Converting it to a blackbody-equivalent emitting area with $F_\nu=A_{\rm bb}B_\nu(T_{\rm hot})/d^2$ gives $A_{\rm bb}=1.5\times10^{19}$--$1.3\times10^{20}\,{\rm m^2}$. If this area is supplied by blackbody-like 1 mm grains with density $3\,{\rm g\,cm^{-3}}$, the corresponding mass is $1.0\times10^{-5}$--$9.0\times10^{-5}\,M_\oplus$. A second estimate comes from the size distribution. A Dohnanyi-like $q=3.5$ cascade normalized to the fitted 0.1--1 $\mu$m surface gives $2.8\times10^{-5}$--$8.9\times10^{-5}\,M_\oplus$ when extended only to millimeter--centimeter grains \citep{Dohnanyi1969,Wyatt2008}. Together with the replenishment requirement for the submicron forsterite surface derived above, these estimates support a reservoir of larger bound parent bodies that supplies both the stable mineral surface and the variable hot-base continuum component.

Material on eccentric orbits could be a returning debris stream or a larger body that periodically crosses the warm reservoir. In the first case, the year-scale infrared recurrence can be understood as the observable response of the hot-base continuum component to repeated delivery of stream material. Each passage of the stream through the periastron or collision-point region can change the dust-production rate, emitting area, or optical depth of the hot-base material while leaving the mineral surface nearly unchanged. This geometry is compatible with simulations in which post-impact debris remains concentrated near collision or periastron regions before becoming progressively more disk-like over tens to hundreds of orbits \citep{Watt2021,Lewis2023,Watt2024}. The infrared recurrence may weaken as collisions, differential precession, and phase mixing proceed, and the returning stream material may become part of a smoother warm debris disk. In the second case, the periodic passage of the larger body could stir pre-existing low-density debris and trigger short-lived collisional cascades. Collisional stirring of the same parent population could also preserve a similar mineral composition. The current light curves and stable mineral emission do not distinguish these mechanisms. Higher-cadence infrared light curves or spectroscopic detection of transient gas features could further constrain these mechanisms.

The combination of spectral stability and quasi-periodic 4.5 $\mu$m excess variability on year timescales therefore moves the observable aftermath of giant impacts beyond a short-lived condensate cloud alone, toward material on eccentric orbits that can sustain warm reservoirs for decades, and potentially longer. Such a process can reduce the need to invoke multiple unrelated impacts to maintain an EDD. Compared with the other 12 debris disks in our sample, ID8 combines a small-grain-dominated surface with pronounced year-scale infrared recurrence. ID8 exposes a transitional route from post-impact ejecta to a sustained warm debris disk.

If material on eccentric orbits is common among EDDs, sparse cadence or low-amplitude variability may already hide similar recurrent behavior. Higher-cadence infrared monitoring over longer baselines can reveal these signals. For ID8, longer-term infrared photometry and occasional mid-infrared spectroscopy could further constrain changes in \mbox{its} recurrence timescale. In addition, SPHEREx spectra can track 4--5 $\mu$m variability in more debris disks and identify candidates for \mbox{JWST}/MIRI spectroscopy. Comparisons of mineral composition across a larger EDD population can then provide statistical constraints on their evolution.

\begingroup
\let\internallinenumbers\relax
\begin{acknowledgments}
We thank the anonymous referee for very insightful and helpful comments. This work was supported by the National Natural Science Foundation of China through projects 12322306, 12373028, 12633007, 12622304 and 12133002. S.W. and X.C. acknowledge support from the Youth Innovation Promotion Association of the CAS (grant Nos. 2023065 and 2022055). This work is based on observations made with the NASA/ESA/CSA James Webb Space Telescope. The data were obtained from the Mikulski Archive for Space Telescopes at the Space Telescope Science Institute. The JWST/MIRI observations of ID8 analyzed here can be accessed via \dataset[doi:10.17909/zvv5-d168]{https://doi.org/10.17909/zvv5-d168}. This work is also based in part on observations made with the \textit{Spitzer Space Telescope}, operated by the Jet Propulsion Laboratory, California Institute of Technology, under a contract with NASA. This publication makes use of data products from the Wide-field Infrared Survey Explorer and NEOWISE, which are projects of the University of California, Los Angeles, and the Jet Propulsion Laboratory/California Institute of Technology, funded by NASA. This publication makes use of data products from the Spectro-Photometer for the History of the Universe, Epoch of Reionization and Ices Explorer (SPHEREx), which is a joint project of the Jet Propulsion Laboratory and the California Institute of Technology, and is funded by the National Aeronautics and Space Administration.
\end{acknowledgments}
\endgroup

\software{Astropy \citep{Astropy2022}, NumPy \citep{Harris2020}, SciPy \citep{Virtanen2020}, Matplotlib \citep{Hunter2007}, DuCK \citep{Kaeufer2024}}

\appendix

\section{State Timing of the Infrared Recurrence}
\label{app:state}

For the state timing, we first combined the data into 30 d bins by inverse-variance weighting within each bin, so that densely sampled sequences did not acquire greater timing weight simply from containing more measurements. We then defined high and low states directly from the binned 4.5 $\mu$m excess curve. High-state bins have $F_j>F_{\rm H}=3.35$ mJy, low-state bins have $F_j<F_{\rm L}=2.15$ mJy, and consecutive selected bins separated by less than 500 d were assigned to the same state. The epoch of each state was defined by a flux-contrast centroid, $t_{\rm H}=\sum_j t_j(F_j-F_{\rm H})/\sum_j(F_j-F_{\rm H})$ and $t_{\rm L}=\sum_j t_j(F_{\rm L}-F_j)/\sum_j(F_{\rm L}-F_j)$, where $t_j$ and $F_j$ are the time and W2-equivalent excess of bin $j$, $F_{\rm H}$ and $F_{\rm L}$ are the high- and low-state thresholds, and the weights are the excess above $F_{\rm H}$ or the deficit below $F_{\rm L}$. Differences between successive centers of the same state define the spacings in Figure~\ref{fig:clock}d. We computed the flux contrast for each recurrence from the paired state-center fluxes. Pairing each of the first five high-state centers with the immediately following low-state center gives high-minus-low contrasts of 2.08, 2.61, 2.11, 2.09, and 2.07 mJy. Their mean and sample standard deviation are $2.19\pm0.23$ mJy, and a linear fit gives a slope of $-0.015\pm0.028$ mJy yr$^{-1}$ ($p=0.62$), consistent with no secular decline. The error bars in Figure~\ref{fig:clock}d are the ranges of same-state spacings recovered when the state-identification procedure is repeated with 15, 30, and 60 d bins, two bin origins separated by half a bin, high-state thresholds of 3.20--3.50 mJy, and low-state thresholds of 2.05--2.25 mJy. The shallow 2013 \textit{Spitzer} [4.5] maximum falls below the adopted high-state threshold. When treated as an additional high state, it gives the open red points in Figure~\ref{fig:clock}d.

\end{document}